\documentstyle[12pt]{article}

 1
\font\twelverm=cmr10 scaled\magstep 1
 1

\font\tenbf=cmbx10
\font\tenrm=cmr10
\font\tenit=cmti10

\font\ninerm=cmr9

\font\eightrm=cmr8

\font\teni=cmmi10   \font\seveni=cmmi7  \font\fivei=cmmi5
\font\tensy=cmsy10
\font\sevenbf=cmbx7
\font\fivebf=cmbx5
\font\sevenrm=cmr7

\font\sevensy=cmsy7
\font\fiverm=cmr5

\font\fivesy=cmsy5
\font\tenex=cmex10
\font\tensl=cmsl10
\font\tentt=cmtt10

\def\tenpoint{\def\rm{\fam0\tenrm}% switch to 10-point type
\textfont0=\tenrm \scriptfont0=\sevenrm \scriptscriptfont0=\fiverm
\textfont1=\teni \scriptfont1=\seveni \scriptscriptfont1=\fivei
\textfont2=\tensy \scriptfont2=\sevensy \scriptscriptfont2=\fivesy
\textfont3=\tenex \scriptfont3=\tenex \scriptscriptfont3=\tenex
\newfam\itfam \newfam\slfam \newfam\ttfam \newfam\bffam
\textfont\itfam=\tenit \def\it{\fam\itfam\tenit}
\textfont\slfam=\tensl \def\sl{\fam\slfam\tensl}
\textfont\ttfam=\tentt \def\tt{\fam\ttfam\tentt}
\textfont\bffam=\tenbf \scriptfont\bffam=\sevenbf
\scriptscriptfont\bffam=\fivebf \def\bf{\fam\bffam\tenbf}
\normalbaselineskip=12pt
\setbox\strutbox=\hbox{\vrule height8.5pt depth3.5pt width0pt}
\let\sc=\eightrm \let\big=\tenbig \normalbaselines\rm
}
\catcode`\@=11
\long\def\@makefntext#1{ %\parindent 1em
\protect\noindent \hbox to 3.2pt {\hskip-.9pt
$^{{\ninerm\@thefnmark}}$\hfil}#1\hfill} %can be used

\def\thefootnote{\fnsymbol{footnote}}
 \def\@makefnmark{\hbox to 0pt{$^{\@thefnmark}$\hss}}  %original

\def\ps@myheadings{\let\@mkboth\@gobbletwo
\def\@oddhead{\hbox{} %\sl
\rightmark\hfil\ninerm\thepage}
\def\@oddfoot{}\def\@evenhead{\ninerm\thepage\hfil %\sl
\leftmark\hbox{}}\def\@evenfoot{}
\def\sectionmark##1{}\def\subsectionmark##1{}}

\begin{document}

%----------------------------PROCSLA.STY---------------------------------------
\newcommand{\symbolfootnote}{\renewcommand{\thefootnote}
	{\fnsymbol{footnote}}}
\renewcommand{\thefootnote}{\fnsymbol{footnote}}
\newcommand{\alphfootnote}
	{\setcounter{footnote}{0}
	 \renewcommand{\thefootnote}{\sevenrm\alph{footnote}}}

%------------------------------------------------------------------------------
%NEW DEFINED SECTION COMMANDS
\newcounter{sectionc}\newcounter{subsectionc}\newcounter{subsubsectionc}
\renewcommand{\section}[1] {\vspace{0.6cm}\addtocounter{sectionc}{1}
\setcounter{subsectionc}{0}\setcounter{subsubsectionc}{0}\noindent
	{\bf\thesectionc. #1}\par\vspace{0.4cm}}
\renewcommand{\subsection}[1] {\vspace{0.6cm}\addtocounter{subsectionc}{1}
	\setcounter{subsubsectionc}{0}\noindent
	{\it\thesectionc.\thesubsectionc. #1}\par\vspace{0.4cm}}
\renewcommand{\subsubsection}[1]
{\vspace{0.6cm}\addtocounter{subsubsectionc}{1}
	\noindent {\rm\thesectionc.\thesubsectionc.\thesubsubsectionc.
	#1}\par\vspace{0.4cm}}
\newcommand{\nonumsection}[1] {\vspace{0.6cm}\noindent{\bf #1}
	\par\vspace{0.4cm}}

%NEW MACRO TO HANDLE APPENDICES
\newcounter{appendixc}
\newcounter{subappendixc}[appendixc]
\newcounter{subsubappendixc}[subappendixc]
\renewcommand{\thesubappendixc}{\Alph{appendixc}.\arabic{subappendixc}}
\renewcommand{\thesubsubappendixc}
	{\Alph{appendixc}.\arabic{subappendixc}.\arabic{subsubappendixc}}

\renewcommand{\appendix}[1] {\vspace{0.6cm}
        \refstepcounter{appendixc}
        \setcounter{figure}{0}
        \setcounter{table}{0}
        \setcounter{equation}{0}
        \renewcommand{\thefigure}{\Alph{appendixc}.\arabic{figure}}
        \renewcommand{\thetable}{\Alph{appendixc}.\arabic{table}}
        \renewcommand{\theappendixc}{\Alph{appendixc}}
        \renewcommand{\theequation}{\Alph{appendixc}.\arabic{equation}}
%       \noindent{\bf Appendix \theappendixc. #1}\par\vspace{0.4cm}}
        \noindent{\bf Appendix \theappendixc #1}\par\vspace{0.4cm}}
\newcommand{\subappendix}[1] {\vspace{0.6cm}
        \refstepcounter{subappendixc}
        \noindent{\bf Appendix \thesubappendixc. #1}\par\vspace{0.4cm}}
\newcommand{\subsubappendix}[1] {\vspace{0.6cm}
        \refstepcounter{subsubappendixc}
        \noindent{\it Appendix \thesubsubappendixc. #1}
	\par\vspace{0.4cm}}

%------------------------------------------------------------------------------
%MACRO FOR ABSTRACT BLOCK  Modified 1/8/94. CLee. \tenrm ==> \tenpoint
\def\abstracts#1{{
	\centering{\begin{minipage}{30pc}\tenpoint\baselineskip=12pt\noindent
	\centerline{\tenpoint ABSTRACT}\vspace{0.3cm}
	\parindent=0pt #1
	\end{minipage} }\par}}

%------------------------------------------------------------------------------
%NEW MACRO FOR BIBLIOGRAPHY
\newcommand{\bibit}{\it}
\newcommand{\bibbf}{\bf}
\renewenvironment{thebibliography}[1]
	{\begin{list}{\arabic{enumi}.}
	{\usecounter{enumi}\setlength{\parsep}{0pt}
%1.25cm IS STRICTLY FOR PROCSLA.TEX ONLY
\setlength{\leftmargin 1.25cm}{\rightmargin 0pt}
%0.52cm IS FOR NEW DATA FILES
%\setlength{\leftmargin 0.52cm}{\rightmargin 0pt}
	 \setlength{\itemsep}{0pt} \settowidth
	{\labelwidth}{#1.}\sloppy}}{\end{list}}

%------------------------------------------------------------------------------
%FOLLOWING THREE COMMANDS ARE FOR 'LIST' COMMAND.
\topsep=0in\parsep=0in\itemsep=0in
\parindent=1.5pc

%LIST ENVIRONMENTS
\newcounter{itemlistc}
\newcounter{romanlistc}
\newcounter{alphlistc}
\newcounter{arabiclistc}
\newenvironment{itemlist}
    	{\setcounter{itemlistc}{0}
	 \begin{list}{$\bullet$}
	{\usecounter{itemlistc}
	 \setlength{\parsep}{0pt}
	 \setlength{\itemsep}{0pt}}}{\end{list}}

\newenvironment{romanlist}
	{\setcounter{romanlistc}{0}
	 \begin{list}{$($\roman{romanlistc}$)$}
	{\usecounter{romanlistc}
	 \setlength{\parsep}{0pt}
	 \setlength{\itemsep}{0pt}}}{\end{list}}

\newenvironment{alphlist}
	{\setcounter{alphlistc}{0}
	 \begin{list}{$($\alph{alphlistc}$)$}
	{\usecounter{alphlistc}
	 \setlength{\parsep}{0pt}
	 \setlength{\itemsep}{0pt}}}{\end{list}}

\newenvironment{arabiclist}
	{\setcounter{arabiclistc}{0}
	 \begin{list}{\arabic{arabiclistc}}
	{\usecounter{arabiclistc}
	 \setlength{\parsep}{0pt}
	 \setlength{\itemsep}{0pt}}}{\end{list}}

%------------------------------------------------------------------------------
%FIGURE CAPTION  Modified 1/8/94. CLee. \tenrm ==> \tenpoint
\newcommand{\fcaption}[1]{
        \refstepcounter{figure}
        \setbox\@tempboxa = \hbox{\tenpoint Fig.~\thefigure. #1}
        \ifdim \wd\@tempboxa > 6in
           {\begin{center}
        \parbox{6in}{\tenpoint\baselineskip=12pt Fig.~\thefigure. #1 }
            \end{center}}
        \else
             {\begin{center}
             {\tenpoint Fig.~\thefigure. #1}
              \end{center}}
        \fi}

%TABLE CAPTION  Modified 1/8/94. CLee. \tenrm ==> \tenpoint
\newcommand{\tcaption}[1]{
        \refstepcounter{table}
        \setbox\@tempboxa = \hbox{\tenpoint Table~\thetable. #1}
        \ifdim \wd\@tempboxa > 6in
           {\begin{center}
        \parbox{6in}{\tenpoint\baselineskip=12pt Table~\thetable. #1 }
            \end{center}}
        \else
             {\begin{center}
             {\tenpoint Table~\thetable. #1}
              \end{center}}
        \fi}

%------------------------------------------------------------------------------
%ACKNOWLEDGEMENT: this portion is from John Hershberger
\def\@citex[#1]#2{\if@filesw\immediate\write\@auxout
	{\string\citation{#2}}\fi
\def\@citea{}\@cite{\@for\@citeb:=#2\do
	{\@citea\def\@citea{,}\@ifundefined
	{b@\@citeb}{{\bf ?}\@warning
	{Citation `\@citeb' on page \thepage \space undefined}}
	{\csname b@\@citeb\endcsname}}}{#1}}

\newif\if@cghi
\def\cite{\@cghitrue\@ifnextchar [{\@tempswatrue
	\@citex}{\@tempswafalse\@citex[]}}
\def\citelow{\@cghifalse\@ifnextchar [{\@tempswatrue
	\@citex}{\@tempswafalse\@citex[]}}
\def\@cite#1#2{{$\null^{#1}$\if@tempswa\typeout
	{IJCGA warning: optional citation argument
	ignored: `#2'} \fi}}
\newcommand{\citeup}{\cite}

%------------------------------------------------------------------------------
%FOR FNSYMBOL FOOTNOTE AND ALPH{FOOTNOTE}
\def\fnm#1{$^{\mbox{\scriptsize #1}}$}
\def\fnt#1#2{\footnotetext{\kern-.3em
	{$^{\mbox{\sevenrm #1}}$}{#2}}}

%----------------------START OF DATA FILE------------------------------

\centerline{Invited review for Strangeness and Quark Matter, Krete,
Sep.~1--5 1994 (World Scientific). `arch-ive/9412316'}
\centerline{\tenbf PHYSICS OF STRANGELETS}
%\vspace{0.8cm}
\vspace{0.5cm}
\centerline{\tenrm JES MADSEN}
\baselineskip=13pt
\centerline{\tenit Institute of Physics and Astronomy, University of Aarhus}
\baselineskip=12pt
\centerline{\tenit DK-8000 {\AA}rhus C, Denmark, e-mail: jesm@dfi.aau.dk}
%\vspace{0.9cm}
\vspace{0.5cm}
\abstracts{After a brief introduction to the physics of bulk strange
quark matter (SQM) this review focuses on the properties of
low baryon number strangelets presently searched for in ultra-relativistic
heavy-ion experiments at CERN and Brookhaven. Shell-model calculations
reveal interesting (meta)stability properties in the experimentally accessible
regime. A liquid drop model (Fermi-gas model) is shown to explain the
overall behavior of the mode-filling calculations, leading to a physical
understanding of strangelet properties, which can be generalized to
non-zero temperature and pressure.}

\vglue 0.3cm
\twelverm
\baselineskip=14pt
\section{Introduction}
While nuclei with baryon number close to that of iron ($A=56$) are
normally assumed to be the lowest energy state of hadronic matter
($E/A\approx 930\hbox{\rm MeV}$), it
has long been realized\cite{bodmer},
that systems composed of a confined Fermi-gas of
up, down, and strange quarks could have an even lower energy per baryon,
thus being absolutely stable. The reason is simple: It is experimentally
known, that ordinary nuclei are stable relative to systems of up and
down quarks, but a lower energy per baryon can be reached by introducing
a third Fermi-sea; that of strange quarks. To create a more stable
system, the energy gained must first compensate for the mass of the
strange quark, but since the typical Fermi-energies involved are
$m_{nucleon}/3\approx 310$MeV, this can be the case if the strange quark
is not too massive (the current mass of the strange quark is believed
to be 100--300 MeV).

Estimates like these were published by Witten\cite{witten} in 1984, whereas
more detailed calculations like those shown in Figure 1 were first
performed by Farhi and Jaffe\cite{farhi}, also 10 years ago.
Those papers defined the birth of strange quark matter physics. Since then
significant progress has been made in the study of the physics and
astrophysics of SQM. Much of this work has been theoretical, but
fortunately the experimental situation is now improving in
the search for cosmic ray strangelets, and
for strangelets produced in ultra-relativistic heavy ion collisions at
CERN and Brookhaven. Ref.\cite{madsen91} contains a general overview of
the field and an extensive list of references. For recent reviews on
experiments, see Ref.\cite{kumar} and Shiva Kumars contribution to the
present volume. A recent review on theory and astrophysics is given in
Ref.\cite{madsen93c}

Most SQM calculations
have been made within the MIT-bag model, which describes bulk systems in
terms of 3 parameters: The bag constant, $B$, which is a measure of the
false vacuum energy that confines the quarks ($B$ thus in practice acts
like an external pressure on the bag); $m_s$, the strange quark mass (up
and down quarks are normally assumed to be massless); and $\alpha_s$,
the strong fine-structure constant, which describes one-gluon exchange
interactions.

\vspace{10.5cm}
\tenrm

\baselineskip=12pt

\noindent
Figure 1. Energy per baryon for bulk SQM as a function of bag constant
and strange quark mass. The strong fine-structure constant is set to
zero. Non-zero $\alpha_s$ roughly corresponds to a rescaling to lower
{\tenit B}.

\twelverm
\baselineskip=14pt

\vspace{0.5cm}

Neither of these parameters are known with sufficient accuracy
from experiments, so there is some freedom in the construction of SQM
models. Fits to the ordinary hadron spectrum have been used to constrain
the parameters, but it is not obvious\cite{farhi} that these values are
the relevant ones for larger assemblies of quarks, so normally the
properties of SQM are studies as a function of these 3 parameters. For
most purposes a non-zero $\alpha_s$ can be ``absorbed'' in a reduction
of $B$, so in the following I shall concentrate on $\alpha_s=0$. Here
a lower limit on $B$ ($B^{1/4}>145$MeV) comes from the stability of
ordinary nuclei relative to up-down quark matter (ordinary nuclei do not
spontaneously decay into strangelets since this would require a high
order weak interaction to make sufficient numbers of strange quarks).
As Figure 1 illustrates, a bag constant smaller than $(164\hbox{\rm MeV})^4$
permits stable bulk SQM for sufficiently low $m_s$, whereas the
metastability window relative to a gas of $\Lambda$'s goes to $(195\hbox{\rm
MeV})^4$. Whereas stability would be of significant interest in
connection with the early Universe and in compact objects like neutron
stars (see the contributions by Alcock, Weber, and Heiselberg, and the
review in\cite{madsen93c}), the
metastability window, which is also interesting (but less conspicuous)
for astrophysics, is clearly much larger. Interestingly, this window is
probably within reach
in ultra-relativistic heavy-ion collisions, because strangelets stable on
a weak interaction time scale may have sufficient time to reach the
detectors, even though finite-size effects destabilize strangelets
relative to bulk SQM.
In the following I shall focus on the physical properties of
(meta)stable strangelets, in particular those with baryon number $A<100$.

\section{Shell model}

The most important property of a strangelet is the mass (or energy per baryon).
A calculation involving mode-filling in a spherical MIT-bag was
first performed by Farhi and Jaffe\cite{farhi}, and later by Greiner
{\it et al.}\cite{greiner} A detailed study for
$ud$-systems was done by Vasak, Greiner and Neise\cite{vasak}.
Gilson and Jaffe\cite{gilson} published a thorough investigation of low-mass
strangelets for 4 different combinations of $s$-quark mass and bag
constant with particular emphasis on metastability against strong
decays, and Madsen\cite{madsen94} studied a wider range of parameters
and explained the results in terms of a liquid drop model.

In the MIT bag model\cite{degrand}
non-interacting quarks are confined to a spherical
cavity of radius $R$. They satisfy the free Dirac equation inside the
cavity and obey a boundary condition at the surface, which corresponds
to no current flow across the surface. The bag itself has an energy of
$BV$. In the simplest version the energy (mass) of the system is given by
the sum of the bag energy and the energies of individual quarks,
\begin{equation}
E=\sum_{i=u,d,s}\sum_{\kappa}N_{\kappa, i}
(m_i^2+k_{\kappa ,i}^2)^{1/2}+B4\pi R^3/3 .
\label{bagenergy}
\end{equation}
Here the momentum $k_{\kappa, i}\equiv x_{\kappa, i}/R$, where
$x_{\kappa, i}$ are eigenvalues of the equation
\begin{equation}
f_\kappa (x_{\kappa ,i})={{-x_{\kappa ,i}}\over{(x_{\kappa ,i}^2
+m_i^2R^2)^{1/2}+m_iR}} f_{\kappa -1}(x_{\kappa ,i}).
\end{equation}
$f_\kappa$ are regular Bessel functions of order $\kappa$ ,
\begin{equation}
f_\kappa (x)=\left\{ \begin{array}{ll}
              j_{\kappa} (x) & \kappa\geq 0 \\
              y_{\kappa} (x)=(-1)^{\kappa +1} j_{-\kappa -1}(x)& \kappa <0
              \end{array}
                \right.
\end{equation}
For states with quantum numbers $(j,l)$ $\kappa$ takes the values
$\kappa =\pm (j+{1\over 2} )$ for $l=j\pm{1\over 2}$. For a given quark flavor
each level has a degeneracy of $3(2j+1)$ (the factor 3 from color
degrees of freedom).
For example, the $1S_{1/2}$ ground-state ($j=1/2$, $l=0$, $\kappa =-1$)
for a massless quark corresponds to
solving the equation $\tan x=x/(1-x)$, giving $x=2.04$. The ground state
has a degeneracy of 6.

For massless quarks (finding the equilibrium radius from $\partial
E/\partial R=0$) one gets
\begin{equation}
E= 364.00\hbox{\rm MeV}B_{145}^{1/4}\left(\sum x_{\kappa ,i} \right)^{3/4}
\label{vaseq}
\end{equation}
where the sum is to be taken over all $3A$ quark-levels, and
$B_{145}^{1/4}\equiv B^{1/4}/145\hbox{\rm MeV}$. The
numbers $x_{\kappa ,i}$ for massless quarks
are tabulated by Vasak {\it et al.}\cite{vasak}

For massive quarks the level filling scheme is more cumbersome.
Fixing bag constant and quark-masses,
for each baryon number one must fill up the lowest energy levels
for a choice of radius; then vary the radius until a minimum energy is
found ($\partial E/\partial R =0$). Since levels cross, the order of
levels is changing as a function of $R$.

Results are shown in Figure 2.
One notices that the energy per baryon smoothly
approaches the bulk limit for $A\rightarrow \infty$, whereas the energy
grows significantly for low $A$. For low $s$-quark mass shells
are recognized for $A=6$ (3
colors and 2 spin orientations per flavor), and less conspicuous ones e.g.
for $A=18$, 24, and 42. As $m_s$ increases it becomes
more and more favorable to use $u$ and $d$ rather than $s$-quarks, and
the ``magic numbers'' change; for instance the first closed shell is
seen for $A=4$ rather than 6.

\vskip 12.5cm

\tenrm
\baselineskip=12pt

\noindent
Figure 2. Energy per baryon (in MeV) for strangelets for a bag constant of
(145MeV)$^4$ (the ``most optimistic'' choice allowed in terms of
strangelet stability; cf.~Sec.~1).
Up and down quark masses are set to zero, whereas the
strange quark mass is varied from 50--300~MeV in steps of 50~MeV (lowest curve
corresponds to lowest mass). The figure to the left shows a blow-up
for low baryon numbers. Further details are given in the text.

\twelverm
\baselineskip=14pt

\vfill\break
{}~
\vskip 18.5cm

\tenrm
\baselineskip=12pt

\noindent
Figure 3. Contour plot of the energy per baryon for strangelets (in MeV)
for the same parameters used in Fig. 2, with the strange quark mass
fixed at 100 MeV, and the baryon number fixed at 20. Total charge (equal to the
number of up quarks less the baryon number) is given on the
abscissa, number of strange quarks on the ordinate.
A total of 1891 strangelet states are included.

\twelverm
\baselineskip=14pt
\vfill\break

Equation (\ref{bagenergy}) can be modified by inclusion of Coulomb
energy, zero-point
fluctuation energy, and color-interaction energy. The
zero-point energy is normally included as a phenomenological term of the
form $-Z_0/R$, where fits to light hadron spectra indicate the choice
$Z_0=1.84$. This was used, for instance, by Gilson and Jaffe\cite{gilson}.
Roughly half of this phenomenological term is due to center-of-mass motion,
which can be included more explicitly by substituting
$\left[\left(\sum x_{\kappa ,i} \right)^2-\sum x_{\kappa ,i}^2 \right]^{3/8}$
instead of $\left(\sum x_{\kappa ,i} \right)^{3/4}$ in Eq.~(\ref{vaseq}).
The proper choice of $\alpha_s$ and $Z_0$ is not straightforward. As
discussed by Farhi and Jaffe\cite{farhi} the values are intimately coupled to
$B$ and $m_s$, and it is not obvious that values deduced from bag model
fits to ordinary hadrons are to be preferred. For reasonable parameter
values one sees a significant effect of the zero-point energy
for $A<10$, but the term
quickly becomes negligible for increasing $A$. The reason for this will
be explained in Section 4.

\vskip 9.5cm

\tenrm
\baselineskip=12pt

\noindent
Figure 4. Contour plot of the energy per baryon for strangelets (in MeV)
for the same parameters used in Fig. 3, except that in the right-hand
plot masses of 5 and 10 MeV have been assumed for the up and down
quarks. Only states with energy per baryon below 940 MeV are shown.

\twelverm
\baselineskip=14pt
\vskip 1cm

Mode-filling calculations as discussed so far have focused on finding
the ground state properties of strangelets, i.e. the lowest mass state
for a given baryon number, $A$. Figures 3 and 4 instead fixes $A=20$ and
displays contours of equal $E/A$ for the whole range of possible
strangelets with that baryon number (a total of 1891 states). What is
quite interesting (and worrying) from an experimental point of view is
the close spacing of states around that of minimum energy. More than a
hundred different states are within 10MeV per baryon from the ground
state. The ground state itself has slightly positive charge for the
present choice of parameters, but neutral as well as negatively charged
states are very close in energy. Many of these states will be stable
against strong decays, and since some weak decays are suppressed by
Pauli-blocking, presumably many different configurations could be
sufficiently long-lived to reach the detectors. This may be good from a
production point of view, but also means that (apart from the large
uncertainties inherent in the theoretical model itself) there is not
going to be {\it a single\/} well-defined strangelet signature, but
rather numerous (meta)stable states and many more possible decay modes!

\section{Liquid drop model}

Shell-model calculations are rather tedious. For many applications a
global mass-formula analogous to the liquid drop model for nuclei is of
great use, and it also gives a physical understanding of the general
properties.

An investigation of the strangelet mass-formula within the
MIT bag model was performed by Berger and Jaffe\cite{berger}. They
included Coulomb corrections and surface tension effects stemming from
the depletion in the surface density of states due to the mass of the
strange quark. Both effects were treated as perturbations added to a
bulk solution with the surface contribution derived from a multiple
reflection expansion.

While showing some of the important physics, this approach
(apart from Coulomb corrections) predicts
constant $E/A$ versus $A$ for $m_s\rightarrow 0$ and
$m_s\rightarrow\infty$, in striking contrast to the shell model results.
Recently it was pointed out that another contribution to
the energy, the curvature term, is dominant (and strongly
destabilizing) at baryon numbers below a
hundred\cite{madsen93a,madsen93b,madsen94}.
The following discussion closely follows Madsen\cite{madsen94}.

I will concentrate on systems small enough
($A\ll 10^7$) to justify neglect of electrons.
Strangelets with $A\ll 10^7$ are smaller than the electron
Compton wavelength, and electrons are therefore
mainly localized outside the quark phase\cite{farhi}.
Thus strangelets do not obey a requirement of local charge
neutrality, as was the case for SQM in bulk. This also leads to a small
Coulomb energy, which is rather negligible for the mass-formula (less
than a few MeV per baryon), but which is decisive for the
quark composition and therefore the charge-to-mass
ratio, $Z/A$, of the strangelet. A characteristic of strangelets, which is
perhaps the best experimental signature, is that this
ratio is very small compared to ordinary nuclei. Indeed, for $m_s=0$ the
ground state strangelet has equal numbers of all three quark flavors and
is therefore charge neutral. Whereas Coulomb effects have been
consistently included\cite{madsen93b},
I will leave out those terms in the equations below.

In the ideal Fermi-gas approximation the energy of a system composed of
quark flavors $i$ is given by
\begin{equation}
E=\sum_i(\Omega_i+N_i\mu_i)+BV
\label{Estrangelet2}
\end{equation}
Here $\Omega_i$, $N_i$ and $\mu_i$ denote thermodynamic potentials,
total number of quarks, and chemical potentials, respectively. $B$ is
the bag constant, $V$ is the bag volume.

In the multiple reflection expansion framework of Balian and
Bloch\cite{balian}, the
thermodynamical quantities can be derived from a density of states of
the form
\begin{equation}
{{dN_i}\over{dk}}=6 \left\{ {{k^2V}\over{2\pi^2}}+f_S\left({m_i\over
k}\right)kS+f_C\left({m_i\over k}\right)C+ .... \right\} ,
\end{equation}
where area $S=\oint dS$ ($=4\pi R^2$ for a sphere) and curvature
$C=\oint\left({1\over{R_1}}+{1\over{R_2}}\right) dS$ ($=8\pi R$ for a
sphere). Curvature radii are denoted $R_1$ and $R_2$. For a spherical
system $R_1=R_2=R$. The functions $f_S$ and $f_C$ are given below.

In terms of volume-, surface-,
and curvature-densities, $n_{i,V}$, $n_{i,S}$, and $n_{i,C}$, the
number of quarks of flavor $i$ is
\begin{equation}
N_i=\int_0^{k_{Fi}}{{dN_i}\over{dk}}dk=n_{i,V}V+n_{i,S}S+n_{i,C}C,
\end{equation}
with Fermi momentum $k_{Fi}=(\mu_i^2-m_i^2)^{1/2}=\mu_i(1-\lambda_i^2)^{1/2}$;
$\lambda_i\equiv m_i/\mu_i$.

The corresponding thermodynamic potentials are related by
\begin{equation}
\Omega_i=\Omega_{i,V}V+\Omega_{i,S}S+\Omega_{i,C}C ,
\end{equation}
where $\partial\Omega_i/\partial\mu_i=-N_i$, and
$\partial\Omega_{i,j}/\partial\mu_i=-n_{i,j}$. The universal volume
terms are given by
\begin{equation}
\Omega_{i,V}=-{{\mu_i^4}\over {4\pi^2}}\left( (1-\lambda_i^2)^{1/2}(1-
{5\over 2}\lambda_i^2)
+{3\over 2}\lambda_i^4\ln{{1+(1-\lambda_i^2)^{1/2}}\over\lambda_i}\right),
\end{equation}
\begin{equation}
n_{i,V}={{\mu_i^3}\over{\pi^2}}(1-\lambda_i^2)^{3/2}.
\end{equation}

The surface contribution from massive quarks is derived from
\begin{equation}
f_S\left({m\over k}\right)=
-{1\over{8\pi}} \left\{ 1-{2\over \pi}
\tan^{-1}{k\over m}\right\}
\end{equation}
as\cite{berger}
\begin{eqnarray}
&&\Omega_{i,S}={3\over{4\pi}}\mu_i^3\left[{{(1-\lambda_i^2)}\over 6}
-{{\lambda_i^2(1-\lambda_i )}\over 3}\right.\cr
&&\left. -{1\over{3\pi}}\left(\tan^{-1}\left[
{{(1-\lambda_i^2)^{1/2}}\over\lambda_i}\right]-2\lambda_i (1-\lambda_i^2)^{1/2}
+\lambda_i^3\ln\left[{{1+(1-\lambda_i^2)^{1/2}}\over\lambda_i}\right]\right)
\right]
\end{eqnarray}
\begin{equation}
n_{i,S}=-{3\over{4\pi}}\mu_i^2\left[{{(1-\lambda_i^2)}\over 2}
-{1\over{\pi}}\left(\tan^{-1}\left[
{{(1-\lambda_i^2)^{1/2}}\over\lambda_i}\right]-\lambda_i (1-\lambda_i^2)^{1/2}
\right)\right] .
\end{equation}
For massless quarks $\Omega_{i,S}=n_{i,S}=0$.

\vfill\break
{}~
\vskip 19.5cm

\tenrm
\baselineskip=12pt

\noindent
Figure 5. Shell-model and liquid drop model calculations are compared for
a bag constant of
(165MeV)$^4$ with massless up and down quarks, and
strange quark mass varied in the range 50--350~MeV in steps of 50~MeV.

\twelverm
\baselineskip=14pt
\vfill\break

The curvature terms have not been derived for massive quarks, but as
shown by Madsen\cite{madsen94}, the following {\it Ansatz\/} works:
\begin{equation}
f_C\left({m\over k}\right)=
{1\over{12\pi^2}} \left\{ 1-{3\over 2}{k\over m}\left(
{\pi\over 2}-\tan^{-1}{k\over m}\right) \right\} .
\label{ansat}
\end{equation}
This expression has the right limit for massless quarks
($f_C=-1/24\pi^2$) and for infinite mass, which corresponds to
the Dirichlet boundary conditions studied by Balian and Bloch\cite{balian}
($f_C=1/12\pi^2$). Furthermore, the expression gives perfect
fits to mode-filling calculations (see Figures 5 and 6).
{}From Eq.~(\ref{ansat}) one derives the following thermodynamic
potential and density:
\begin{equation}
\Omega_{i,C}={{\mu_i^2}\over{8\pi^2}}\left[\lambda_i^2\log{{1+
(1-\lambda_i^2)^{1/2}}\over{\lambda_i}} +{\pi\over{2\lambda_i}}-
{{3\pi\lambda_i}\over{2}}+\pi\lambda_i^2-{1\over\lambda_i}\tan^{-1}
{{(1-\lambda_i^2)^{1/2}}\over{\lambda_i}}\right] ;
\end{equation}
\begin{equation}
n_{i,C}={\mu_i\over{8\pi^2}}\left[(1-\lambda_i^2)^{1/2}-
{{3\pi}\over 2}{{(1-\lambda_i^2)}\over\lambda_i}+{3\over\lambda_i}
\tan^{-1}{{(1-\lambda_i^2)^{1/2}}\over\lambda_i}\right] .
\end{equation}
Notice that $\Omega_{i,C}$ and $n_{i,C}$ are non-zero for massless
quarks in contrast to the surface terms. Thus not only $s$-quarks, but
to an even higher degree $u$ and $d$-quarks contribute significantly to
the finite-size effects.

With these prescriptions the differential of $E(V,S,C,N_i)$ is given by
\begin{equation}
dE=\sum_i\left(\Omega_{i,V}dV+\Omega_{i,S}dS+\Omega_{i,C}dC+\mu_idN_i
\right) +BdV .
\label{dE}
\end{equation}

Minimizing the total energy at fixed $N_i$ by taking $dE=0$ for a sphere
gives the pressure equilibrium constraint
\begin{equation}
B=-\sum_i\Omega_{i,V}-{2\over R}\sum_i\Omega_{i,S}
-{2\over R^2}\sum_i\Omega_{i,C}.
\label{bag}
\end{equation}
The ``optimal'' composition is found by minimizing the energy
with respect to $N_i$ at fixed $A$ and radius, giving
\begin{equation}
0=\sum_i\mu_idN_i .
\end{equation}
Together with Eq.~(\ref{Estrangelet2}) these constraints give the
properties of a spherical quark lump in its ground state.

The solution is compared with the shell model calculations in Figures 5
and 6.
The fits are remarkably good, indicating that all the important physics
(apart from the shells) can be understood in terms of the surface and
curvature contributions.

\section{Bulk approximation}

Simpler, approximate mass-formulae can be derived by using a bulk
approximation to the chemical potentials. Including bulk terms only,
the energy minimization, Eq.~(\ref{bag}) (with
$\lambda\equiv\lambda_s$, and superscript $^0$ denoting bulk values),
changes to
\begin{eqnarray}
B&=&-\sum_i\Omega_{i,V}^0\cr
&=&\sum_{i=u,d}{{(\mu_i^0)^4}\over{4\pi^2}}
+ {{(\mu_s^0)^4}\over{4\pi^2}}\left[
 (1-\lambda^2)^{1/2}(1-
{5\over 2}\lambda^2)
+{3\over 2}\lambda^4\ln{{1+(1-\lambda^2)^{1/2}}\over\lambda}\right],
\end{eqnarray}
and the baryon number density is now given by
\begin{equation}
n_A^0={1\over 3}\left[\sum_{i=u,d}{{(\mu_i^0)^3}\over{\pi^2}}+
{{(\mu_s^0)^3}\over{\pi^2}}(1-\lambda^2)^{3/2}\right] .
\end{equation}
A bulk radius per baryon is defined by
\begin{equation}
R^0\equiv (3/4\pi n_A^0)^{1/3}.
\end{equation}

In bulk equilibrium the chemical potentials of the three quark flavors
are equal, $\mu_u^0=\mu_d^0=\mu_s^0\equiv\mu^0=\epsilon^0/3$, where
$\epsilon^0$ is the bulk energy per baryon.
One may approximate the energy per baryon of small
strangelets as a sum of bulk, surface and curvature terms, using the
chemical potential calculated in bulk:
\begin{equation}
\epsilon\equiv {E\over A}=\epsilon^0 +A^{-1}\sum_i\Omega_{i,S}^0 S^0
+A^{-1}\sum_i\Omega_{i,C}^0 C^0 ,
\end{equation}
where $S^0=4\pi (R^0)^2A^{2/3}$ and $C^0=8\pi (R^0)A^{1/3}$.
Examples for $B^{1/4}=145\hbox{\rm MeV}$ are (with $s$-quark mass
given in parenthesis; all energies in MeV)
\begin{eqnarray}
\epsilon (0)&=&829+0A^{-1/3}+351A^{-2/3}\cr
\epsilon (50)&=&835+61A^{-1/3}+277A^{-2/3}\cr
\epsilon (100)&=&852+83A^{-1/3}+241A^{-2/3}\cr
\epsilon (150)&=&874+77A^{-1/3}+232A^{-2/3}\cr
\epsilon (200)&=&896+53A^{-1/3}+242A^{-2/3}\cr
\epsilon (250)&=&911+22A^{-1/3}+266A^{-2/3}\cr
\epsilon (300)&=&917+0A^{-1/3}+295A^{-2/3} .
\end{eqnarray}

The bulk approximations above are generally good to better than 2MeV for
$A>100$, 5MeV for $A\approx 50$, 10MeV for $A\approx 10$ and 20MeV for
$A\approx 5$, when compared to the perfect fit of the liquid drop model
(see Figure 6).
They always undershoot relative to the correct solution. This is because
the actual chemical potentials of the quarks increase
when $A$ decreases, whereas the bulk approximations use constant
$\mu$. For massless $s$-quarks the expression for $\epsilon (0)$ scales
simply as $B^{1/4}$. The same scaling applies for $m_s >\epsilon^0/3$,
where no $s$-quarks are present; in the example above the scaling can be
applied to $\epsilon (300)$. For intermediate $s$-quark masses no simple
scaling applies. For instance, if $B^{1/4}=165$MeV one finds $\epsilon
(150) = 985+93A^{-1/3}+265A^{-2/3}$;
$\epsilon (250) = 1027+46A^{-1/3}+284A^{-2/3}$.
Coulomb effects were not included above. Their
inclusion would have no influence for $m_s\rightarrow 0$, but would
change the results by a few MeV for large $m_s$.

In connection with the shell-model calculations I described the effects
of a zero-point energy of the form $-Z_0/R$, and claimed that it was
important only for $A<10$. This can be understood in the
bulk approximation of constant $\mu$, because the zero-point term per
baryon is proportional to $A^{-4/3}$ compared to $A^{-1/3}$ and $A^{-2/3}$ for
surface and curvature energies. The full term to be added to the bulk
approximation expressions for a given $\epsilon^0$ is:
\begin{equation}
\epsilon_{zero}=-Z_0(4/243\pi )^{1/3}\left[ 2+[1-(3m_s/\epsilon^0)]^{3/2}
\right] ^{1/3}\epsilon^0 A^{-4/3} ,
\end{equation}
typically of order $-200 Z_0\hbox{\rm MeV} A^{-4/3}$.

Writing $E/A =\epsilon_0 + c_{surf}A^{-1/3}+c_{curv}A^{-2/3}$, with
$c_{surf}\approx 100$MeV and $c_{curv}\approx 300$MeV (the Coulomb
energy is negligible in comparison, because strangelets are almost
neutral), the stability condition $E/A<m_n$ may be
written as $A>A_{min}^{abs}$, where
\begin{equation}
A_{min}^{abs}= \left(
{{c_{surf}+[c_{surf}^2+4c_{curv}(m_n-\epsilon_0)]^{1/2}}
\over{2(m_n-\epsilon_0)}}\right)^3 .
\end{equation}
Stability at baryon number 30 requires a bulk binding energy in excess
of 65 MeV, which is barely within reach for
$m_s > 100\hbox{\rm MeV}$ if, at the same time,
$ud$-quark matter shall be unstable (see Figure 1).
The proposed cosmic ray strangelet-candidates with baryon number
370\cite{saito} would for
stability require a bulk binding energy per
baryon exceeding 20~MeV
to overcome the combined curvature and surface energies.
Absolute stability relative to a gas of $^{56}$Fe corresponds to furthermore
using 930~MeV instead of $m_n$, whereas stability relative to a gas
of $\Lambda$-particles (the ultimate limit for
formation of short-lived strangelets)
would correspond to substitution of $m_\Lambda = 1116$MeV.

One can also calculate the minimum baryon
number for which long-lived metastability is possible.
Identifying this as the limit of neutron
emission requires
$dE_{curv}/dA +dE_{surf}/dA < m_n-\epsilon_0$, or
\begin{equation}
A_{min}^{meta}= \left(
{{c_{surf}+[c_{surf}^2+3c_{curv}(m_n-\epsilon_0)]^{1/2}}
\over{3(m_n-\epsilon_0)}}\right)^3 .
\end{equation}
To have $A_{min}^{meta}<30$ requires $m_n-\epsilon_0>30$MeV, which is
possible, but only for a narrow range of parameters.

However, it is important to notice, that
shell effects can have a stabilizing effect.
As stressed by Gilson and Jaffe\cite{gilson} the fact that the slope of $E/A$
versus $A$ becomes very steep near magic numbers can render
strangelets that are metastable (stable against single baryon emission)
even for $\epsilon^0 > 930$MeV.  Also,
the time-scales for energetically allowed decays have not been calculated.
The existence of small baryon number strangelets is ultimately an
experimental issue.
\vfill\break
{}~
\vskip 19cm

\tenrm
\baselineskip=12pt

\noindent
Figure 6. Shell-model and liquid drop model calculations are compared for
a bag constant of
(145MeV)$^4$ with massless up and down quarks, and
strange quark mass varied in the range 50--300~MeV in steps of 50~MeV.
For each set
of calculations the lower smooth curve is the bulk approximation to the
liquid drop model, whereas the upper smooth curve is the full liquid
drop model.

\twelverm
\baselineskip=14pt
\vfill\break

\section{Lessons for experiments}
The lessons for experimental strangelet searches can be summarized as
follows:
\begin{itemize}
\item[1)]
(Meta)stable strangelets are possible.
\item[2)]
Energy per baryon, charge etc are strongly parameter dependent (though
normally $|Z|\ll A$).
\item[3)]
A liquid drop model (Fermi-gas model) explains the general properties.
Curvature is a decisive effect.
\item[4)]
Strangelets have prominent shell structure for low $A$.
\item[5)]
Many isotopes are almost degenerate in energy.
\item[6)]
As a consequence of 5) the decay modes for metastable strangelets are
numerous. Lifetimes are not predictable at present, but could well be
larger (due to Pauli-blocking) than those of the strange hadronic matter
discussed by St{\"o}cker in these Proceedings, thus providing a possible
experimental distinction (also, strange hadronic matter can not be
absolutely stable in contrast to strangelets).
\end{itemize}

The conclusions above are mainly based on studies within the simplest
version of the MIT bag model. Effects of zero-point energy and finite
$\alpha_s$ have only been included in a crude fashion, and actual QCD
calculations are beyond reach. Real strangelets
can have non-spherical shape. They are created at a high temperature
(which tends to increase the energy per baryon). Et cetera.

Clearly our theoretical understanding of strangelets is still in a
rather primitive state. What we really need before making more
complicated models is experimental data!

\section{Acknowledgements}
I thank Apostolos Panagiotou and Georges Vassiliadis for organizing such
a wonderful week for us in Krete, and Shiva Kumar, Charles Alcock,
Horst St\"ocker, and Dan M{\o}nster Jensen for providing useful comments.
This work was supported by
the Theoretical Astrophysics Center
under the Danish National Research Foundation.

\end{document}